\documentclass[aps,prl,showpacs,twocolumn,amsmath,amssymb]{revtex4-1}
\usepackage{graphicx,bbm}

\newcommand{\be}{\begin{equation}}
\newcommand{\ee}{\end{equation}}

\newcommand{\bra}[1]{{\langle #1 \vert}}

\newcommand{\ket}[1]{{\vert #1 \rangle}}

\newcommand{\braket}[2]{\langle #1 \vert #2 \rangle}

\newcommand{\ave}[1]{{\langle #1\rangle}}
\newcommand{\ii}{ {\rm i} }
\newcommand{\dd}{ {\rm d} }
\newcommand{\ZZ}{\mathbb{Z}}

\newcommand{\CC}{\mathbb{C}}

\newcommand{\z}{{\rm z}}

\newcommand{\mm}[1]{{\mathbf{#1}}}

\def\tr{{{\rm tr}}}

\def\one{\mathbbm{1}}

\newcommand{\La}{{\mathtt L}}
\newcommand{\Ra}{{\mathtt R}}

\begin{document}

\title{Families of quasi-local  conservation laws and quantum spin transport}
\author{Toma\v{z} Prosen and Enej Ilievski}
\affiliation{Department of Physics, FMF,  University of Ljubljana, Jadranska 19, 1000 Ljubljana, Slovenia}

%\date{June 15 2011, Accepted to Physical Review Letters on August 12, 2011}

\date{\today}

\begin{abstract}
For fundamental integrable quantum chains with deformed symmetries we outline a general procedure for defining a continuous family of quasi-local operators whose  time-derivative is supported near the two boundary sites only. The program is implemented for a spin 1/2 $XXZ$ chain, resulting in improved rigorous estimates for the high temperature spin Drude weight. 
\end{abstract}

 
\maketitle

{\em Introduction.-} Quantum dynamics of locally interacting systems in one-dimension \cite{giamarchi} continues to pose fundamental challenges to theorists. For example, complete integrability \cite{kbi93}, which implies existence of a macroscopic number (i.e. of the same order as the number of particles or degrees of freedom) of local conservation laws can prevent a system from thermalizaing \cite{rigol,barthel} or stop a current from decaying \cite{zotos,affleck,s12},  but precise conditions for when and mechanisms for how this can happen are still generally unclear.
For instance, it is possible that due to general symmetry arguments all the local conserved quantities following from quantum inverse scattering method \cite{kbi93} are irrelevant for the interesting physical observables under study, like magnetization or spin current. This happens for example in the anisotropic Heisenberg spin 1/2 chain, the so-called $XXZ$ model, and allows for a surprising suggestion \cite{robin,pz} of spin diffusion in the Ising-like regime of high spin-coupling anisotropy $|\Delta|>1$.  Nevertheless, for $|\Delta|\le 1$, numerical \cite{fabian,karrasch} and experimental \cite{solugbenko,qsimul} evidence exists for anomalous, or even ballistic transport which is characterized by positivity of spin Drude weight, despite the fact that at zero magnetization (or in the absence of external magnetic field) the spin current is orthogonal to all local conserved operators \cite{gm95} and so the Mazur bound \cite{mazur} for the Drude weight vanishes \cite{zotos}.

The problem has recently been resolved by finding a `missing conservation law' $Z$ \cite{p11a}, namely it has been shown that a quasi-local operator exists in the form of a rapidly converging series of local operators which almost commutes with the Hamiltonian in the sense that the residual terms are supported only near the boundary of the chain and Drude weight can still be bounded away from zero rigorously \cite{ip13}. However this exotic new object $Z$ appeared rather mysterious as the technique for deriving it \cite{p11a} did not seem connected to integrability structures such as transfer-operators and Yang-Baxter equations (YBE), neither it gave any clue on how it may be generalized to other models. In this Letter we answer these puzzles by deriving a whole family of quasi-local conservation laws as a function of complex parameter. These new objects, including $Z$ of Ref.~\cite{p11a} as a special case, are derived from a novel, so-called {\em highest-weight} quantum Yang-Baxter transfer operator based on an infinitely dimensional complex spin $s$ representation of the quantum group $U_q(\mathfrak{sl}_2)$, which is the symmetry of the $XXZ$ model. Quasi-locality here emerges as a consequence of differentiation with respect to $s$ at the trivial point $s=0$, unlike in the standard case \cite{kbi93,gm95} where locality is a consequence of taking the logarithm of the {\em trace} transfer operator in the fundamental representation $s=1/2$. 
We show how the new continuous family of conservation laws can be applied to yield improved rigorous Mazur bound \cite{ip13} on spin Drude weight.
We focus our analysis to the case of $XXZ$ spin 1/2 chain, but it should be generalizable to other fundamental integrable models sharing quantum group Yang-Baxter structure.

{\em Holomorphic family of almost-conserved quasi-local operators.-}
The starting point is in acknowledging (see e.g. \cite{k01,k02,ttf83}) a general Yang-Baxter equation (YBE) in a triple vector space ${\cal V}_{s_1} \otimes {\cal V}_{s_2} \otimes {\cal V}_{s_3}$
where $s_1,s_2,s_3 \in \CC$ denote arbitrary complex representation parameters for generally infinitely dimensional representations of quantum group $U_q(\mathfrak{sl}(2))$. 
These so-called Verma modules ${\cal V}_s$, spanned by a semi-infinite orthonormal (ON) basis $\ket{k},k\in\ZZ^+$, are generated by deformed spin-$s$ operators 
%$\mm{S}^\alpha_s \in {\rm End}({\cal V}_s)$
\begin{eqnarray}
\mm{S}^\z_s &=& \sum_{k=0}^\infty (s-k) \ket{k}\bra{k}, \nonumber\\
\mm{S}^+_s &=& \sum_{k=0}^\infty \frac{\sin(k+1)\lambda}{\sin\lambda} \ket{k}\bra{k+1}, \label{verma} \\
\mm{S}^-_s &=& \sum_{k=0}^\infty \frac{\sin(2s-k)\lambda}{\sin\lambda} \ket{k+1}\bra{k}, \nonumber
\end{eqnarray}
satisfying the quantum group relations $[\mm{S}^+_s,\mm{S}^-_s]=\sin[2\lambda \mm{S}^\z_s]/\sin\lambda$, $[\mm{S}^\z_s,\mm{S}^\pm_s]=\pm \mm{S}^\pm_s$. For clarity of notation we use bold symbols to denote objects which are not scalars over 
infinitely dimensional module ${\cal V}_s$.  For $s\in\ZZ^+/2$, ${\cal V}_s$ is reducible to a finite, $2s+1$ dimensional irrep. The deformation parameter $q=\exp(\ii\lambda)$ is related to the
anisotropy parameter $\Delta=\cos\lambda$ of the $n$-spin 1/2 $XXZ$ Heisenberg chain with Hamiltonian
\begin{eqnarray}
H_n&=&\sum_{x=1}^{n-1} \one_{2^{x-1}} \otimes h \otimes \one_{2^{n-x-1}},\cr
h&=& 2\sigma^+ \otimes \sigma^- + 2\sigma^-\otimes \sigma^+ + \Delta \sigma^\z \otimes \sigma^\z,
\end{eqnarray}
which can be considered as an operator over ${\cal V}_{1/2}^{\otimes n}\simeq \CC^{2^n}$ and $\sigma^{\pm,\z},\sigma^0=\one_2$ is a set of standard Pauli matrices.

Let us define a two-parametric Lax-operator in terms of the universal $R-$matrix over ${\cal V}_{s}\otimes{\cal V}_{1/2}$ \cite{k02}, i.e. a $2\times 2$ matrix with entries in ${\rm End}({\cal V}_s)$
\begin{equation}
\mm{L}(\varphi,s) = \begin{pmatrix}
\sin(\varphi+\lambda \mm{S}^\z_s) & (\sin\lambda) \mm{S}^-_s \cr
(\sin\lambda) \mm{S}^+_s & \sin(\varphi-\lambda \mm{S}^\z_s) \cr
\end{pmatrix}.
\end{equation}
Then the YBE in ${\cal V}_{s}\otimes {\cal V}_{s'}\otimes {\cal V}_{1/2}$ together with the fact that $\bra{0}\otimes\bra{0}$ ($\ket{0}\otimes\ket{0}$) is a left (right) eigenvector of the $R$-matrix over ${\cal V}_{s}\otimes {\cal V}_{s'}$
guarantees commutativity of the {\em highest weight transfer-operators} \cite{pip13}
\begin{equation}
W_n(\varphi,s) = \bra{0} \mm{L}(\varphi,s)^{\otimes n} \ket{0}.
\end{equation}
Namely, for any pair of {\em spectral} parameters $\varphi,\varphi'\in\CC$ and {\em representation} parameters $s,s'\in\CC$, we have
\begin{equation}
[W_n(\varphi,s),W_n(\varphi',s')] = 0.
\end{equation}
Note that the special case $S=W_n(\pi/2,s)$ is exactly the Cholesky factor of the nonequilibrium steady state density operator $S S^\dagger$ \cite{p11b,kps} of the boundary driven $XXZ$ chain, with Lindblad jump operators $L_1 = \sqrt{\varepsilon}\sigma^- \otimes \one_{2^{n-1}}$, $L_2 = \sqrt{\varepsilon}\one_{2^{n-1}} \otimes \sigma^+$, if $\cot(s\lambda) = \varepsilon/(2\ii\sin\lambda)$.

The operators $W_n(\varphi,s)$ are in general non-local and are not commuting with the Hamiltonian $H_n$, however setting spectral parameters to $\varphi'=\varphi+\delta$ in the YBE and expanding to first order in $\delta$ results in a fundamental {\em divergence relation} for local two-site commutators \cite{S70,sklyanin}
\begin{equation}
[h,\mm{L} \otimes\mm{L}] = 2\sin\lambda (\mm{L} \otimes \mm{L}_\varphi - \mm{L}_\varphi\otimes\mm{L} ),
\label{divergence}
\end{equation}
where $\mm{L}\equiv\mm{L}(\varphi,s)$, $\mm{L}_\varphi\equiv \partial_\varphi\mm{L}(\varphi,s) = \cos\varphi \cos(\lambda\mm{S}^\z_s)\otimes\sigma^0-\sin\varphi\sin(\lambda\mm{S}^\z_s)\otimes\sigma^\z$.

Of fundamental importance to gain quasi-locality in these objects is a derivation with respect to a complex (deformed spin) representation parameter at $s=0$, which is implied by the following observation:
\medskip

\noindent
{\bf Lemma:}
{\em 
Let us define a modified auxiliary space $\tilde{\cal V}$ with a split vacuum, namely $\ket{0}$ replaced by a pair of distinct highest weight states $\ket{\La}$ and $\ket{\Ra}$, i.e.  $\tilde{\cal V}$ being spanned by a formal ON basis $\{\ket{\La},\ket{\Ra},\ket{1},\ket{2},\ldots\}$.
Let $\tilde{\mm{S}}^\alpha=\mm{S}^\alpha_0|_{\tilde{\cal V}}$ denote projected spin operators, which are essentially given by (\ref{verma}) with summation index $k$ running from $1$ \cite{note}, and define a modified Lax matrix
\begin{equation}
\tilde{\mm{L}}(\varphi) = \sum_{\alpha\in\{0,\pm,\z\}} \tilde{\mm{L}}^\alpha(\varphi) \otimes \sigma^\alpha
\end{equation}
with components
\begin{eqnarray*}
\tilde{\mm{L}}^0(\varphi) &=& \ket{\La}\bra{\La} +  \ket{\Ra}\bra{\Ra}  + \cos(\lambda \tilde{\mm{S}}^\z),  \nonumber \\
\tilde{\mm{L}}^\z(\varphi)&=& \cot\varphi\sin(\lambda \tilde{\mm{S}}^\z), \label{Lcomp} \\
\tilde{\mm{L}}^+(\varphi) &=& \ket{1}\bra{\Ra} + \frac{\sin\lambda}{\sin\varphi} \tilde{\mm{S}}^-, \quad \tilde{\mm{L}}^-(\varphi) =  \ket{\La}\bra{1} +\frac{\sin\lambda}{\sin\varphi} \tilde{\mm{S}}^+.  \nonumber
\end{eqnarray*}
Consequently, we define also the corresponding modified heighest weight transfer operator
\begin{equation}
Z_n(\varphi) = \bra{\La} \tilde{\mm{L}}(\varphi)^{\otimes n}\ket{\Ra}.
\label{Zn}
\end{equation}
Then, the normalized $s$-derivative at $s=0$ can be expressed as
\begin{equation}
\frac{1}{(\sin\varphi)^{n}}\partial_s W_n(\varphi,s)\vert_{s=0} =
\frac{2\lambda}{\sin\lambda} Z_n(\varphi) +  \lambda\cot\varphi M^\z_n,
\label{lemma}
\end{equation}
where $M^\z_n=\sum_{x=1}^n \one_{2^{x-1}}\otimes \sigma^\z \otimes \one_{2^{n-x}}$ is the total magnetization operator.}

The proof is just a formal expression of the fact that at $s=0$ the transitions $\ket{0}\to\ket{0}$ in $\partial_s \bra{0}\mm{L}(\varphi,s)^{\otimes n}\ket{0}|_{s=0}$, expressed via Leibniz rule applied over $n-$fold matrix product operator, are
only possible (i) via virtual states $\ket{1},\ket{2}\ldots$ if $s$-derivative `acts' on the amplitude $\bra{1}\mm{S}^-_s\ket{0}$
(\ref{verma}), which otherwise would vanish as $s=0$, or (ii) directly where $s$-derivative acts on the amplitude at $\bra{0}\mm{S}^\z_s\ket{0}$. The cases (i,ii) correspond to the first, second term on the RHS of Eq.~(\ref{lemma}), respectively. Note that all the operators under discussion commute with $M^\z_n$, $[W_n(\varphi,s),M^\z_n]=[H_n,M^\z_n]=0$, hence also $Z_n(\varphi)$ form a commuting family
\begin{equation}
[Z_n(\varphi),Z_n(\varphi')] = 0,\quad \forall \varphi,\varphi'\in\CC.
\end{equation}

We note that Eq.~(\ref{Zn}) generates a translationally invariant matrix product operator (MPO)
\begin{equation}
Z_n =\!\!\!\!\!\!\! \sum_{\alpha_1,\alpha_2\ldots,\alpha_n}\!\!\!\!\bra{\La} \tilde{\mm{L}}^{\alpha_1} \tilde{\mm{L}}^{\alpha_2}\cdots\tilde{\mm{L}}^{\alpha_n}\ket{\Ra}\sigma^{\alpha_1}\otimes\sigma^{\alpha_2}\otimes\cdots\sigma^{\alpha_n},
\label{MPO}
\end{equation}
which can be written as a sum of local terms, since
 $\bra{\La}\tilde{\mm{L}}^\alpha = \delta_{\alpha,0}\bra{\La}+\delta_{\alpha,-}\bra{1}$, $\tilde{\mm{L}}^\alpha\ket{\Ra}=\delta_{\alpha,0}\ket{\Ra}+\delta_{\alpha,+}\ket{1}$, 
\begin{equation}
Z_n = \sum_{r=2}^n \sum_{x=0}^{n-r} \one_{2^{x}} \otimes q_r \otimes \one_{2^{n-r-x}}.
\label{qm}
\end{equation}
where $q_r$ is an $r-$site density, i.e. an element of ${\rm End}({\cal V}_{1/2}^{\otimes r})$ which acts non-trivially on sites $1$ and $r$. In other words, $q_r$ is of exact MPO form (\ref{MPO}) for $n=r$ with $\alpha_1=-,\alpha_r=+$.
We define \cite{ip13} the operator $Z\equiv Z_{n=\infty}$ of the infinite chain to be {\em quasi-local} if $\exists \gamma,\xi > 0$ such that $\|q_r\| < \gamma \exp(-\xi r)$, and we call operator sequence $Z_n$ to be {\em almost conserved} if for any $n$,
$[H_n,Z_n]=\sum_{r=1}^n (b_r \otimes \one_{2^{n-r}} - \one_{2^{n-r}}\otimes b_r)$, where $b_r \in {\rm End}({\cal V}_{1/2}^{\otimes r})$ and $\exists \gamma',\xi' > 0$ such that $\|b_r\| < \gamma' \exp(-\xi' r)$.
We have shown in previous work \cite{p11a,ip13} that existence of quasi-local almost-conserved operators (QLAC) implies non-trivial bounds on ballistic transport.

We are now in position to state the main result:
\medskip

\noindent
{\bf Theorem:}
{\em 
For a dense set of commensurate easy-plane anisotropies $\lambda = \pi l/m$, $l,m\in \ZZ$, $l \le m > 0$,
the operators $Z(\varphi)$ are strictly quasi-local and almost-conserved for all $\varphi \in {\cal D}_{m} \subset\CC$ where ${\cal D}_m=\{\varphi; |{\rm Re}\, \varphi-\frac{\pi}{2}| < \frac{\pi}{2m}\}$ is an open vertical strip of width $\pi/m$ centered around $\varphi_0=\pi/2$. Furthermore, $Z(\varphi)$ is holomorphic on ${\cal D}_m$.}
\smallskip

\noindent
{\bf Proof:} We start by tensor-multiplying the local divergence relation (\ref{divergence}) by $\mm{L}^{\otimes(j-1)} \otimes \bullet \otimes \mm{L}^{\otimes(n-j-1)}$, then summing over $j=1\ldots n-1$, taking the highest weight state expectation value $\bra{0}\bullet\ket{0}$, and finally differentiating $\partial_s\bullet|_{s=0}$.
Using the Lemma (\ref{lemma}) and carefully book-keeping all the terms, we finally arive at the key identity
\begin{eqnarray}
&& [H_n,Z_n(\varphi)] =  \sigma^\z \otimes \one_{2^{n-1}} - \one_{2^{n-1}}\otimes\sigma^\z \nonumber\\
 &&- 2\sin\lambda\cot\varphi \left(\sigma^0 \otimes Z_{n-1}(\varphi) - Z_{n-1}(\varphi)\otimes \sigma^0\right).\;\;
\label{HZ}
\end{eqnarray}
Exploring one-element-per-row property of the matrices $\tilde{\mm{L}}^\alpha$, the Hilbert-Schmidt product $(A,B):=2^{-n}\tr(A^\dagger B)$ of any pair of $Z_n(\varphi)$,
can be calculated in terms of a two-parametric transfer matrix \cite{note2}
\begin{eqnarray}
&& K_n(\varphi,\varphi') := (Z_n(\overline{\varphi}),Z_n(\varphi')) = \bra{\La}\mm{T}(\varphi,\varphi')^n\ket{\Ra}, \label{KT} \\
&& \mm{T}(\varphi,\varphi') :=  \ket{\La}\bra{\La} +  \ket{\Ra}\bra{\Ra} + \frac{\ket{\La}\bra{1}}{2}+ \frac{\ket{1}\bra{\Ra}}{2} + \mm{T}'(\varphi,\varphi')\nonumber \\
&&  \mm{T}'(\varphi,\varphi'):=\sum_{k=1}^\infty\Bigl\{ (\cos^2(k\lambda)+\cot\varphi\cot\varphi'\sin^2(k\lambda))\ket{k}\bra{k} \nonumber  \\
&&\quad+\frac{|\sin(k\lambda)\sin(k+1)\lambda|}{2\sin\varphi\sin\varphi'}\left(\ket{k}\bra{k+1} + \ket{k+1}\bra{k}\right)\Bigr\}. \label{Tp}
\end{eqnarray}
For $\lambda=\pi l/m$, the transition $\ket{m}\to\ket{m+1}$ is forbidden, $\bra{m}\mm{T}\ket{m+1}=0$, so $\mm{T}$ can be replaced by a $(m+1)\times (m+1)$ matrix truncated to a finite set of states $\ket{\La},\ket{\Ra},\ket{1},\ldots,\ket{m-1}$ with a {\em symmetric tridiagonal} matrix $\mm{T}'$ being its orthogonal projection to the last $m-1$ states. Then we prove a sequence of statements: (i) $\mm{T}'$ is strictly {\em contracting}, i.e. for all its eigenvalues $\tau_j,j=1\ldots,m-1$, sorted as $|\tau_1|>|\tau_2|>\ldots$, we have $|\tau_j| < 1$ if $\varphi,\varphi'\in{\cal D}_m$. First, let us assume $\varphi'=\overline{\varphi}$ and write ${\rm Re\,}\varphi=\frac{\pi}{2}+u$.
Defining a positive diagonal matrix $\mm{D}=\sum_{k=1}^{m-1}\left|\sin\frac{\pi l k}{m}\right| \ket{k}\bra{k}$, 
and tridiagonal Toeplitz matrix $\mm{A}=\sum_{k=1}^{m-1} \cos(2u)\ket{k}\bra{k}-\frac{1}{2}\sum_{k=1}^{m-2}(\ket{k}\bra{k+1}+\ket{k+1}\bra{k})$, we have
$\one - \mm{T}' = |\sin\varphi|^{-2}\mm{D}\mm{A}\mm{D}$. 
Matrix elements of $\mm{T}'$ are real and non-negative so leading eigenvalue should be positive $\tau_1>0$, and $\mm{T}'$ is contracting if $\one - \mm{T}' > 0$. This is equivalent to condition $\mm{A} > 0$ which holds if $|u| < \frac{\pi}{2m}$, i.e. $\varphi \in {\cal D}_m$.
For general $\varphi,\varphi'\in{\cal D}_m$, $\mm{T}'$ is still contracting as a consequence of Cauchy-Schwartz inequality $|K_n(\varphi,\varphi')|^2 \le 
K_n(\overline{\varphi},\varphi) K_n(\overline{\varphi'},\varphi')$.
(ii) $\tau_j$ are also eigenvalues of $\mm{T}$, whereas the eigenvectors $\ket{\tau_j}'$ of $\mm{T}'$  map to the corresponding eigenvectors of $\mm{T}$ via $\ket{\tau_j}=\ket{\tau_j}' + \ket{\La}\frac{\braket{1}{\tau_j}'}{2\tau_j-2}$.
(iii) Furthermore, $\mm{T}$ has an eigenvalue $\tau_0=1$ of multiplicity $2$ with a single eigenvector $\ket{\tau_0}=\ket{\La}$ and a defective eigenvector $\ket{\psi}=\psi_{\Ra}\ket{\Ra}+\sum_j \psi_j \ket{j}$, $(\mm{T}-\one)\ket{\psi_0} = \ket{\tau_0}$ whose components can be calculated from bottom-up substitution using explicit form (\ref{KT}), resulting in a recurrence $\psi_{m-k}=[T_{m-k,m-k-1}/(1-T_{m-k,m-k})] C^{-1}_{k-1} \psi_{m-k-1}$ and $\psi_1=2$, $\psi_{\Ra}=2[T_{21}/(1-T_{22})]C_{m-2}^{-1}$ where $C_k$ form a continued fraction sequence $C_0=1$, $C_{k+1}=1-1/[4\cos^2(\varphi+\varphi')C_k]$.  Implementing Jordan decomposition of Ref.~\cite{p11a} one finally obtains an explicit estimation
$K_n(\varphi,\varphi') = n K(\varphi,\varphi') + {\cal O}\left(\tau_1^{n}\right)$ where 
\begin{equation}
K(\varphi,\varphi') = -\frac{\sin\varphi \sin\varphi'}{2\sin^2(\pi l/m)} \frac{\sin\left((m-1)(\varphi+\varphi')\right)}{\sin\left(m(\varphi+\varphi')\right)}.
\label{Kt}
\end{equation}
Note that $K(\varphi,\varphi')$ is non-singular when $\varphi,\varphi' \in {\cal D}_m$, whereas $K(\overline{\varphi},\varphi) = \lim_{n\to\infty} (Z_n(\varphi),Z_n(\varphi))/n$ is becoming singular exactly for 
${\rm Re}\,\varphi = \pi/2 \pm \pi/(2m)$, i.e. on $\partial {\cal D}_m$. For densities $q_r$ (\ref{qm}) we write 
$(q_r(\varphi),q_r(\varphi)) = \bra{1}\mm{T}'(\overline{\varphi},\varphi)^r\ket{1}$, following (\ref{Tp}), implying together with elementary operator-norm inequality $\|A\|^2 \le (A,A)$:
\begin{equation}
\|q_r(\varphi)\| \le \gamma \left|\tau_1(\overline{\varphi},\varphi)\right|^{r/2}, \quad {\rm for\; some}\quad \gamma > 0.
\end{equation}
This proves quasi-locality of $Z(\varphi)$ for $\varphi\in{\cal D}_m$ with exponent $\xi=-\frac{1}{2}\log|\tau_1|>0$.
Almost-conservation with the same exponent $\xi'=\xi$ follows by rewriting the second line of Eq.~(\ref{HZ}) as
$Z_{n-1}(\varphi)\otimes \sigma^0-\sigma^0 \otimes Z_{n-1}(\varphi)=\sum_{r=2}^n (q_r \otimes \one_{2^{n-r}} - \one_{2^{n-r}}\otimes q_r)$. 
$Z(\varphi)$ is also holomorphic on ${\cal D}_m$ as it is given in terms of exponentially converging sum, in operator norm, of strictly local operators, each of which is holomorphic in $\varphi$. QED

We note that $Z_n(\pi/2)$ is exactly an isolated QLAC $Z^\dagger$ constructed in Ref.~\cite{p11a} via alternative model-specific method, whereas technique described here should be readily generalizable to other integrable models with deformed symmetries.

{\em Integral form of Mazur bound for spin Drude weight.-} Here we will show how the continuous family of QLAC $Z_n(\varphi)$ can be applied 
to rigorously estimate the spin Drude weight $D$ which yields the ballistic contribution to the real part of spin conductivity $\sigma'(\omega)=2\pi D \delta(\omega) + \sigma'_{\rm reg}(\omega)$. Within the linear response theory the Drude weight can be expressed in terms of time-correlation function as $D=\lim_{t\to\infty}\lim_{n\to\infty}\frac{\beta}{2nt}\int_0^t \dd t'\ave{J_n(t')J_n}_\beta$, where
$J_n(t) = e^{\ii t H_n}J_n e^{-\ii t H_n}$, $\ave{\bullet}_\beta = \tr(e^{-\beta H_n}\bullet)/\tr e^{-\beta H_n}$, and $J_n =\sum_{x=1}^{n-1} \one_{2^{x-1}} \otimes j \otimes \one_{2^{n-x-1}}$ is the spin-current with density $j = \ii \sigma^+ \otimes \sigma^- - \ii\sigma^-\otimes \sigma^+$.

\begin{figure}
        \centering	
         \vspace{-3mm}
	\includegraphics[width=0.95\columnwidth]{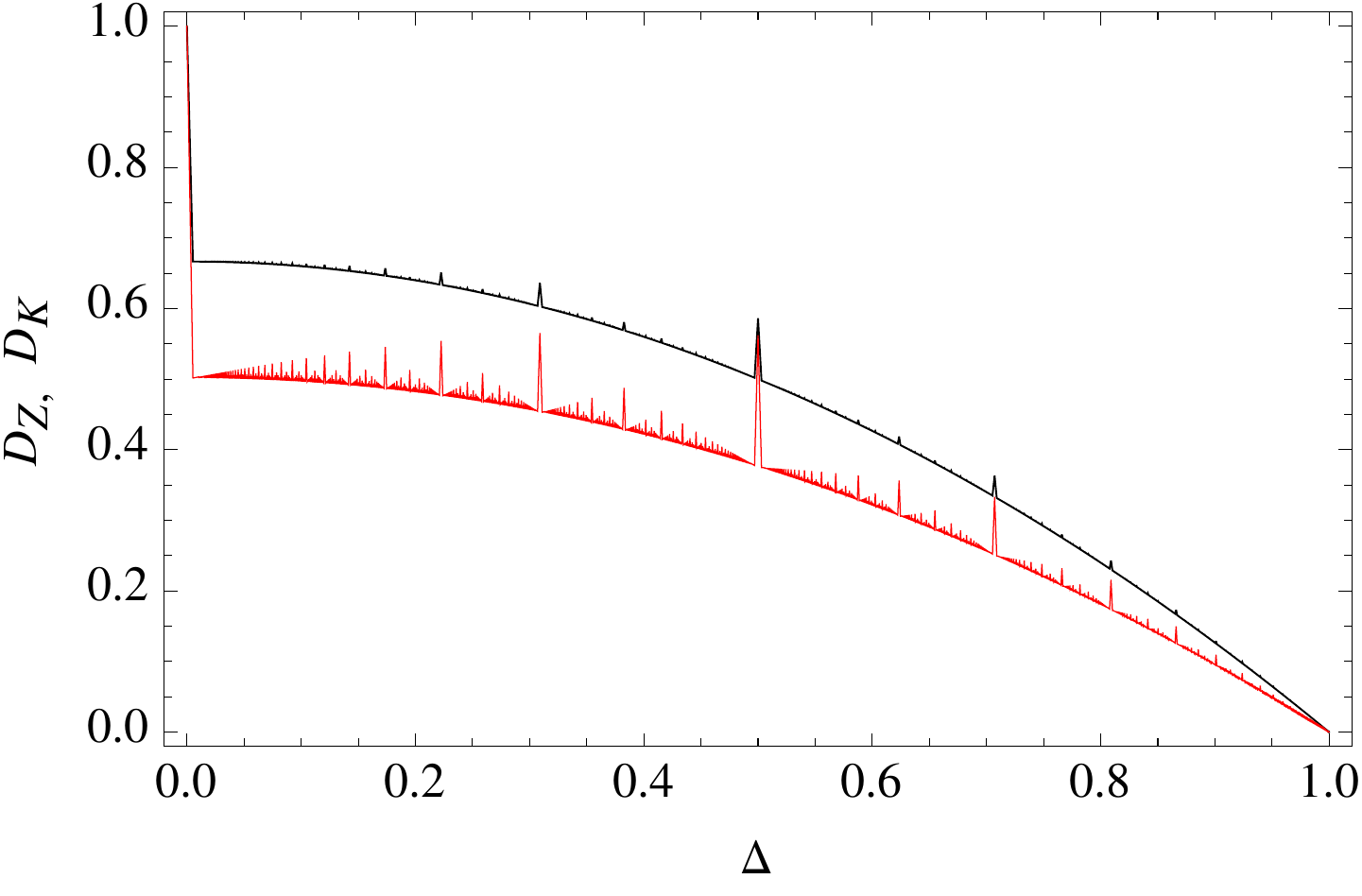}
	\vspace{-3mm}
	\caption{(Color online) Optimized Mazur bound $D_K$ (\ref{DK}) (black) versus the bound $D_Z$ of Ref.~\cite{p11a} (red) which is based on a single quasi-local almost-conserved operator $Z_n(\pi/2)$.}
	\label{fig:DK}
\end{figure}

Limiting ourselves, for simplicity, to infinite temperature $\beta=0$ we apply the rigorous form of the Mazur bound \cite{mazur,zotos}, namely Theorem 2 of Ref.~\cite{ip13}, stating that 
$D \ge \lim_{n\to\infty}\frac{\beta}{2n}\sum_{k,l}(J_n,Q_k)(U^{-1})_{k,l}(Q_l,J_n)$ where $U_{k,l}=(Q_k,Q_l)$ is a positive definite matrix and $\{ Q_k\}$ is an arbitrary set of linearly-independent QLACs (noting that they need not be Hermitian). Here we take an incountable continuum of them, namely $\{Z_n(\varphi)\}\cup\{Z^\dagger_n(\varphi)\}$ labelled by points $\varphi$ from a two-dimensional analyticity strip $\varphi\in{\cal D}_m$. Using  elementary identities $(J_n,Z_n(\varphi))=
-(J_n,Z^\dagger_n(\varphi))\equiv -\frac{\ii(n-1)}{4}$, and $(Z_n(\varphi),Z^\dagger_n(\varphi'))\equiv 0$, we arrive at the Drude weight estimate $D \ge \frac{\beta}{4}D_K$ with
\begin{equation}
D_K = \frac{1}{4} \int_{{\cal D}_m}\!\!\!\dd^2 \varphi f(\varphi),
\end{equation}
where $f(\varphi)$ is the solution of the complex-plane Fredholm integral equation of the first kind 
\begin{equation}
\int_{{\cal D}_m}\!\!\!\dd^2\varphi' K(\varphi,\varphi')f(\varphi') = 1, \quad \varphi \in {\cal D}_m.
\end{equation}
The kernel $K(\varphi,\varphi')$ defines a positive definite operator, substituting for the matrix $\frac{1}{n}U_{k,l}$ in \cite{ip13}, which we essentially have to invert.
Fortunately, the form of solution can be guessed in our case (\ref{Kt}), namely $f(\varphi) = c/|\sin\varphi|^4$ where $c$ is a constant which can be determined by elementary integration, 
yielding an explicit, closed form expression for the Drude weight bound (see also Fig.~\ref{fig:DK})
\begin{equation}
D_K = \frac{\sin^2(\pi l/m)}{\sin^2(\pi/m)}\left(1 - \frac{m}{2\pi}\sin\left(\frac{2\pi}{m}\right)\right).
\label{DK}
\end{equation}
This is a non-trivial improvement over the previous lower bound $D_Z = \frac{m}{2(m-1)}\sin^2\left(\frac{\pi l}{m}\right)  = \frac{m}{2(m-1)}(1-\Delta^2)$ \cite{p11a} based on a single QLAC $Z_n(\pi/2)$, $D_K > D_Z$, but again is a nowhere differentiable function of $\Delta$ and, remarkably, agrees with one of the debatable Bethe ansatz results \cite{zotos2} at $\lambda = \pi/m$. It seems we have now fully explored the known Yang-Baxter structure of the problem hence we dare to conjecture that our bound (\ref{DK}) should in fact be saturated. One might suggest that higher $s$-derivatives $(\dd/\dd s)^k W(\varphi,s)|_{s=0}$ could also be candidates for further independent QLACs, however a brief inspection shows that already the second derivative $k=2$ at $\varphi=\pi/2$ is a non-local operator.

{\em Discussion.-}
We have outlined a procedure for derivation of families of quasi-local conservation laws  of $XXZ$ chain which are orthogonal to previously known \cite{gm95} strictly local conserved quantities.
The latter are given, for periodic boundary conditions, in terms of logarithmic $\varphi$-derivatives of trace of monodromy matrix in fundamental representation $F_n^{(k)} = (\dd/\dd \varphi)^k \log \tr \mm{L}(\varphi,\frac{1}{2})^{\otimes n}|_{\varphi=\lambda/2}$ and are irrelevant for the spin transport in the absence of external magnetic fields since $(J_n,F_n^{(k)})=0$. The former, however, can be derived using related though more involved integrability concepts, namely in terms of derivation of a highest-weight (vacuum) diagonal element of quantum monodromy matrix with respect to complex spin representation parameter at $s=0$. Interestingly enough, our concept can be mapped to a logarithmic $s$-derivative at arbitrary value of $s\in\CC$ by a suitable symmetrized shift of the spectral parameter, namely
\begin{equation}
\partial_s \log \tilde{W}_n(\varphi,s) = \partial_p \frac{\tilde{W}_n(\varphi+s\lambda,p)+\tilde{W}_n(\varphi-s\lambda,p)}{2}\vert_{p=0},
\label{log}
\end{equation}
where $\tilde{W}_n(\varphi,s) := W_n(\varphi,s)/\sin^n(\varphi+s\lambda)$ is a normalized highest weight transfer operator satisfying another interesting property
\begin{equation}
\tilde{W}_n(\varphi,-s)\tilde{W}_n(\varphi,s) = \one.
\label{ident}
\end{equation}
While (\ref{ident}) is easy to prove straightforwardly by writing the product on LHS in ${\cal V}_{-s}\otimes{\cal V}_s$, the relation (\ref{log}) remains a conjecture based on extensive empirical evidence for 
finite chains.

We thank I. Affleck for pointing out how to construct defective eigenvectors of matrices of type (\ref{KT}) and
acknowledge support by Slovenian ARRS grant  P1-0044.


\begin{thebibliography}{10}

\bibitem{giamarchi} T. Giamarchi, ``Quantum physics in one dimension'', (Clarendon, Oxford, 2004).

\bibitem{kbi93} V.~E. Korepin, N.~M. Bogoliubov and A.~G. Izergin, ``Quantum inverse scattering method and correlation functions'', (Cambridge Univ. Press, Cambridge 1993).

\bibitem{rigol} M.~Rigol, V.~Dunjko and M.~Olshanii, Nature {\bf 452}, 854 (2008).

\bibitem{barthel} T.~Barthel and U.~Schollw\" ock Phys. Rev. Lett. {\bf 100}, 100601 (2008).

\bibitem{zotos} X. Zotos, F. Naef and P. Prelov\v sek, Phys. Rev. B {\bf 55}, 11029 (1997).

\bibitem{affleck} J.~Sirker, R.~G.~Pereira and I.~Affleck, Phys. Rev. Lett. {\bf 103}, 216602 (2009); Phys. Rev. B {\bf 83}, 035115 (2011).

\bibitem{s12} J.~Sirker, Int. J. Mod. Phys. B {\bf 26}, 1244009 (2012).

\bibitem{robin}  R.~Steinigeweg and J.~Gemmer, Phys. Rev. B. {\bf 80}, 184402 (2009); R.~Steinigeweg, Phys. Rev. E {\bf 84}, 011136 (2011).

\bibitem{pz} T. Prosen and M. \v Znidari\v c, J. Stat. Mech., P02035 (2009); M.~\v Znidari\v c, Phys. Rev. Lett. {\bf 106}, 220601 (2011).

\bibitem{fabian} F. Heidrich-Meisner et al., Eur. Phys. J. Special Topics {\bf 151}, 135 (2007).  

\bibitem{karrasch} C.~Karrasch, J.~H.~Bardarson and J.~E.~Moore, Phys. Rev. Lett. {\bf 108}, 227206 (2012); C.~Karrasch {\em et al.} {\tt  arXiv:1301.6401} (2013).

\bibitem{solugbenko} A. V. Sologubenko et al., J. Low Temp. Phys. {\bf 147}, 387 (2007).

\bibitem{qsimul} J. Simon {\em et al.}, Nature {\bf 472}, 307 (2011).

\bibitem{gm95} M. P. Grabowski and P. Mathieu,  Ann. Phys. (N.Y.) {\bf 243}, 299 (1995).

\bibitem{mazur} P. Mazur, Physica {\bf 43}, 533 (1969).

\bibitem{p11a} T. Prosen, Phys. Rev. Lett. {\bf 106}, 217206 (2011).

\bibitem{ip13} E. Ilievski and T. Prosen, Commun. Math. Phys. {\bf 318}, 809 (2013).

\bibitem{k01} S.~E.~Derkachov, D.~Karakhanyan and R.~Kirschner, Nucl. Phys. B {\bf 618}, 589 (2001).

\bibitem{k02} D.~Karakhanyan, R.~Kirschner and M.~Mirumyan,  Nucl. Phys. B {\bf 636}, 529 (2002).

\bibitem{ttf83} V.~O.~Tarasov, L.~A.~Takhtajan, L.~D.~Faddeev, Teor. Mat. Fiz. {\bf 57}, 1059 (1983).

\bibitem{pip13} T.~Prosen, E.~Ilievski and V.~Popkov,  {\tt  arXiv:1304.7944}.

\bibitem{p11b} T. Prosen, Phys. Rev. Lett. {\bf 107}, 137201 (2011).

\bibitem{kps} D. Karevski, V. Popkov and G. M. Sch\" utz,  Phys. Rev. Lett. {\bf 110}, 047201 (2013).

\bibitem{S70}  B.~Sutherland, J. Math. Phys. {\bf 11}, 3183 (1970).

\bibitem{sklyanin} E.~K.~Sklyanin, `Quantum Inverse Scattering Method. Selected Topics', {\tt arXiv:hep-th/9211111} (1992). 

\bibitem{note} Note that such truncated spin operators $\tilde{\mm{S}}^\alpha$ in fact generate ${\cal V}_{-1}$, while ${\cal V}_0$ is one-dimensional.

\bibitem{note2} We note \cite{p11a} that $\mm{T}:=\sum_\alpha \frac{1}{2}\tr((\sigma^\alpha)^\dagger \sigma^\alpha) \mm{\tilde{L}}^\alpha(\varphi)\otimes\mm{\tilde{L}}^\alpha(\varphi')$
preserves the diagonal subspace, identifying $\ket{k}\otimes\ket{k}\to\ket{k}$. In addition, we apply diagonal similarity transformation which makes the transfer matrix $\mm{T}'$ (\ref{Tp}) symmetrtic.

\bibitem{zotos2} X. Zotos, Phys. Rev. Lett. {\bf 82}, 1764 (1999); J. Benz {\em et al.}, J. Phys. Soc. Jpn. Supp. {\bf 74},  181 (2005).

\end{thebibliography}
\end{document}